\documentclass[11pt,a4paper]{article}
\usepackage{graphicx}
\usepackage[left=2cm,right=2cm,top=2cm,bottom=2cm]{geometry}
\begin{document}
\title{Soft modes in jammed hard spheres (I): \\
Mean field theory of the isostatic transition }
\author{Giorgio Parisi\\Dipartimento di Fisica,
 Universit\`a di Roma {\it La Sapienza},
\\
INFN, Sezione di Roma I, IPCF -- CNR\\
P.le A. Moro 2, I-00185 Roma, Italy
}
\date{}
\maketitle
\abstract{In this paper we  consider different models for soft modes in jammed hard spheres. We show how one can construct mean field models that can be solved analytically. The analytic solution of these models displays an excess of low energy soft modes that become more and more localized by decreasing the energy. A simple solution of these models is found in the infinite dimensional limit.}

\section{Introduction}
In the last years  an excess of low energy modes in models of granular materials and of compressed soft spheres  has been discovered \cite{REV0,REV}. These modes are related to many crucial properties of jammed material and of low temperature glasses (e.g. the Boson peak). Our aim is to obtain detailed predictions for realistic models. However, following an old tradition in statistical mechanics, here we start  by studying mean field models.

In this paper we want to understand the origin of this phenomenon in mean field models. This excess of zero models in granular materials is clearly related to the condition of isostaticity \cite{REV0,REV}: i.e. the average the number  of contacts of  each sphere $z_i$ is equal to twice the dimension of the space ($\overline{z}=2D$), i.e. it is just barely minimum needed to implement mechanical stability. We will try to put this phenomenon under a more general setting: at this end  we will  study a few models that can be solved analytically.

In order to understand the origin of the excess of low energy modes it is natural to put forward the following conjecture \cite{DD}.  We consider a Hamiltonian with $N$ degrees of freedom (i.e. an $N \times N$ matrix). We suppose that the corresponding quadratic form can be written as \begin{equation}
Q(\delta)\equiv \sum_{i,k=1,N}J_{i,k}\delta_i\delta_k=\sum_{s=1,M} Q^s(\delta)\ , 
\end{equation}
where $Q^k(\delta)$ are $M$ {\sl simple}\footnote{The quadratic forms $Q^k(\delta)$ are different from zero only if $\delta$ belongs to a low dimensional random subspace.} non-negative random quadratic forms and $\delta$ is an $N$ dimensional vector. The simplest case is given by the Wishart matrices (discussed later) where $ Q^s(\delta)=\delta \cdot x^s$, $x^s$ being a random vector.

We are interested to the thermodynamic limit with $\alpha\equiv M/N$ constant. It is clear that $Q(\delta)$ is non-negative definite, however the density of zero model may be different from zero. 
We suppose that the fraction of zero modes $\zeta(\alpha)$ is non-zero at $\alpha<\alpha_c$, it is zero for  $\alpha\geq\alpha_c$ and it vanishes linearly (due to geometric considerations) at $\alpha_c$. 

Following some of the ideas exposed in \cite{DD}, we conjecture that in the generic case the following phenomena happen:
\begin{itemize}
\item  When  $\alpha \neq\alpha_c$    the spectral density $\rho(\lambda)$ has a gap (or a quasi-gap \footnote{In some case the spectral density may be not strictly equal to zero, but it could be exponentially small. In this paper we not investigate the behaviour of the density in these tails.}), i.e. the function $\rho(\lambda)$ is zero (or extremely small) the interval $[0:\lambda_c(\alpha)]$  ($\lambda_c(\alpha)\ne 0$).
\item At $\alpha=\alpha_c$ the quantity $\lambda_c(\alpha_c)$ is zero and  the spectral density $\rho(\lambda)$  is proportional to $\lambda^{-1/2}$. If we define the {\sl frequency} $\omega \equiv\lambda^{1/2}$, the spectral density in $\omega$ (i.e. $D(\omega)=2 \omega \rho (\omega^2)$), is flat around $\omega=0$.
\item Only at $\alpha=\alpha_c$ \footnote{At $\alpha>\alpha_c$  there is an effective band-edge at positive $\omega$. It is likely that the states become strictly localised near the band edge, but we have not studied this point.} there is an avoided localization transition at $\omega=0$. More precisely, let us use the matrix notation and let us consider
the resolvent 
\begin{equation}
R(\lambda)={1\over \lambda-J}\ .
\end{equation}

The function 
\begin{equation}
\rho(\lambda)={Tr(R(\lambda))\over N}
\end{equation} is an analytic function $\lambda$ that has a cut on the real axis in the thermodynamic limit $N\to\infty$. The diagonal matrix elements of $R$  usually play an important role in the theory.

We can focus our attention of the quantity $I(\lambda)$ defined for $\lambda$ real as 
\begin{equation}
I(\lambda)= \lim_{\epsilon \to 0}\lim_{N \to \infty}{\overline {\left(\left( Im(R_{i,i}(\lambda+i\epsilon)\right)^2\right)} \over \left (\overline {Im (R_{i,i}(\lambda+i\epsilon))}\right) ^2} \label{I}\, ,
\end{equation}
where the bar denotes the average over the sites $i$. 

In the localized phase the quantity $I(\lambda)$ is divergent  and the r.h.s of eq. (\ref{I}) behaves as $IPR/\epsilon$, where $IPR$ is the inverse participation ratio. When we stay in the extended phase $I(\lambda)$ is finite: a divergence of $I(\lambda)$ is usually a signal of a localization transition. 

We will argue that $I(\lambda)$ is divergent when  $\lambda$ goes to zero in the same way if there where a localization transition\footnote{There is actually no localisation transition because at negative $\lambda$ the spectral density is zero.} at $\lambda=0$. Near $\lambda=0$ we should see quasi-localized extended states and this is likely related to the results of \cite{QUASI}.
\end{itemize}

In this paper we will consider some models that have been already introduced in the literature \cite{DD}. However in most of the cases the models have been studied by explicit diagonalization of $N\times N$ matrices or by approximate	methods \cite{EFFECTIVE};  here we are interested to derive results in the limit $N\to\infty$  by exact analytic computations.

In section 2 we present the simplest possible model where we can observe  the appearance of a singularity  in the distribution of soft modes: i.e. Wishart matrices where the spectral density is given by the well-known Marcenko-Pastur distribution.

Sections 3 and 4 are the core of the paper:  we apply the methods developed for soft spheres
 in \cite{3,5} to the case of harmonic jammed spheres (in  papers  \cite{3,5} the reader can find a detailed discussion of the method we use). We find that also in these mean field models for  oscillations of harmonic spheres the spectral density has a square root divergence at small $\lambda$ when the isostaticity condition is satisfied. 

In section 5 we show that in the random spring model that we have studied in the previous two sections, the eigenstates have a tendency to become more and more localized at low energies. Finally in the section 6 we show how this model can be simply solved analytically in the infinite dimensional limit.

\section{A simple instructive model}

Let  consider the following model (pseudo-Wishart matrices).
We construct the $N\times N$ matrix $J$ as
\begin{equation}
J_{i,k}= \sum_{\nu=1,M}x^\nu_i x^\nu_k \, , \ \ Q(\delta)=\sum_{\nu=1,M} \left( x^\nu \cdot \delta\right)^2 \, ,
\end{equation}
where the $x$'s are Gaussian distributed vectors of dimension $N$: we  normalized  the $x$'s to 1, but this is irrelevant in the $N \to \infty$ limit.  The {\sl bona fide} Wishart matrices  have Gaussian $x$'s, here we set $|x|=1$. The two cases coincide in the $N \to \infty$, but for finite $N$ they are different.

The spectral properties of Wishart matrix are very well know and they have been computed in many different way \cite{BP}. For completeness we sketch a derivation  based on replicas taken from \cite{BP1}. We are interested to the case where both $M$ and $N$ go to infinity at constant $\alpha=M/N$.

We must follow exactly the same steps of the computation done for the memory capacity of the Hopfield model \cite{AMIT}. The main difference with the approach of Amit et al.  \cite{AMIT} is that in our case the spins are Gaussian (in their case the spins are $\pm 1$). 

We use  the replica formalism and we introduce two $n \times n$ matrices $R$ and $Q$ (eventually we have to send $n$ to 0). Using the saddle point method (that is correct for large $N$) we arrive at the equations from the replicas matrices. One finds a solution where both  matrices $R$ and $Q$ are diagonal. We denote the values of the diagonal elements by $r$ and $q$. Their  physical meaning can be found by looking at there values at the saddle point: they satisfy the following relations:
\begin{equation}
q(\lambda)={\sum_{i=1,N} G_{i,i}\over N} \equiv g(\lambda)\, , \ \ \ \ r(\lambda)={\sum_{\nu=1,M} \sum_{i,k=1,N} x^\nu_i x^\nu_k G_{i,k}\over M}\, ,
\end{equation}
and therefore $q(\lambda)$ is equal to the trace of the resolvent $\rho(\lambda)$.
 
If we repeat the same steps of \cite{AMIT}, we find
\begin{equation}
r=1/(1-q) \ \, , \, q=1/(\lambda-\alpha r)  \ \  \to \ \ \ q=1/(\lambda -\alpha/(1-q)) \label{CavityW} \, .
\end{equation}
We finally  get
\begin{equation}
\lambda q^2-(\lambda-\alpha+1)q+1=0 \  \to \ q(\lambda,\alpha)={\lambda-\alpha+1-\sqrt{(\lambda-\alpha+1)^2-4 \lambda}\over 2 \lambda} \, .
\end{equation}
Near $\lambda=0$ we get
\begin{equation}
q(\lambda,\alpha)={-\alpha+1+|\alpha-1|\over 2  \lambda}+O(1) \, .
\end{equation}
Apart from the singularity that is present at $\lambda=0$, the spectrum extends in the interval
\begin{equation}
\alpha+1\pm2\sqrt{\alpha} \, .
\end{equation}
Indeed the spectral density is given by the well-known Marcenko-Pastur distribution:
\begin{equation}
\rho(\lambda)=\theta(1- \alpha) (1- \alpha)\delta(\lambda)+Re\left({\sqrt{4 \lambda -(\lambda-\alpha+1)^2}\over 2 \lambda} \, .\right)
\end{equation}

 It is easy to see  the properties of the spectrum directly from the equation (\ref{CavityW}) from a small $\lambda$ expansion without using the explicit solution of second-degree equations. Indeed at $\alpha=1$ the equation for $q$  can be simplified in the region of large $q$ and small $\lambda$. One finds that  equation (\ref{CavityW})  reduces to 
 \begin{equation}
q=q-1-\lambda q^2 \ \  \to \ \ \ q(\lambda) \approx (-\lambda)^{-1/2} \, .
\end{equation}
   The mechanism (i.e. the cancellation of the leading term in $q$ between the r.h.s. and the l.h.s) gives naturally a $1/\sqrt{\lambda}$ behavior for the resolvent and it will be present also in the other cases we will consider later. This kind of behaviour implies that $D(\omega)$ has a non-zero limit when $\omega$ goes to zero.

It should be possible to derive eq. (\ref{CavityW}) using a cavity approach as explained in  \cite{MPV}. The equation for $q$ is the cavity equation for going from $N$ to $N+1$, while the equation for $r$ is the cavity equation for going from $M$ to $M+1$.

\section{A  mean field model of random springs}
 We consider  a model for small oscillations of harmonic compressed spheres. The quadratic part  of the energy of small oscillations  is given by the following  quadratic form \begin{equation}
Q(\delta)=\frac12\sum_{\{i,k\}\in {\cal C}} \left((\delta_i-\delta_k)\cdot x_{i,k}\right)^2 \equiv \sum_{i,k} \delta_i \cdot J_{i,k}\cdot\delta_k \, .
\label{QUA}
\end{equation}
where the sum runs over the pairs of spheres in contact: both the $\delta$'s and the $x_{i,k}$'s ($x_{i,k}=-x_{k,i}$) are vectors in a $D$ dimensional space. 

Usually the lattice (or network) of spheres and the corresponding vectors  $x_{i,k}$ are obtained by compressing a gas of hard or harmonic spheres and the vectors $x_{i,k}$ are given by $x_i-x_k$, where $x_i$ is the position of the $i^{th}$ particle.  The average coordination number of the lattice ($z$)is given by $2 M/N$ where $M$ the number of contact pairs. If we neglect rattlers, the average coordination number satisfies the isostatic condition $z=2D$ in the limit of small pressure.

It is clear that we cannon compute the spectrum of $J$ in a simple form. However it has been noted in the framework of the computation of normal modes \cite{3,5} of oscillations of soft spheres that we can associate to $J$ a scrambled matrix $J'$ defined as follows \cite{3}: to each of the unordered pair $\{i,k\}$ we associate a random new pair
 $\{i'(i,k),k'(i,k)\}$ such that the new quadratic form is given by:
\begin{equation}
Q'(\delta)=\frac12\sum_{\{i,k\}\in {\cal C}} ((\delta_{i'(i,k)}-\delta_{k'(i,k)})\cdot x_{i'(i,k),k'(i,k)})^2 \, .\label{QUI}
\end{equation}
The spectrum of $J'$ is computable analytically. This procedure has been follow in \cite{3,5} for non-harmonic soft spheres.
The final lattice is a Bethe lattice with a Poisson  distribution of the local coordination number ($z_i$) and the $x$'s are just random vectors.  The model is very similar to the one recently  been studied by Liu and Manning (the DD model in their notation) \cite{DD} and we agree with their conclusions on the spectrum.

Unfortunately a Poisson distribution of the local coordination number ($z_i$) clearly produces artifacts in the spectrum:   isolated particles  are present with finite probability $\exp(-z)$ and these isolated particles  contribute to zero modes (like rattlers). Moreover, if the original problem is mechanically stable,  all the $z_i$ are greater than $D$. We could avoid these problem by constraining  also the final distribution to have $z_i>D$ (e.g. we could set $z_i=D+1+ \tilde z_i$, where $\tilde z_i$ is a Poisson variable with average $z-4$).  

Here we simplify the whole analysis by setting $z_i=z$ (obviously $z$ is an integer.) We finally arrive to a Bethe lattice of fixed co-ordination $z$ and $N$ points.

\begin{figure}\begin{center}
\includegraphics[width=.7\textwidth]{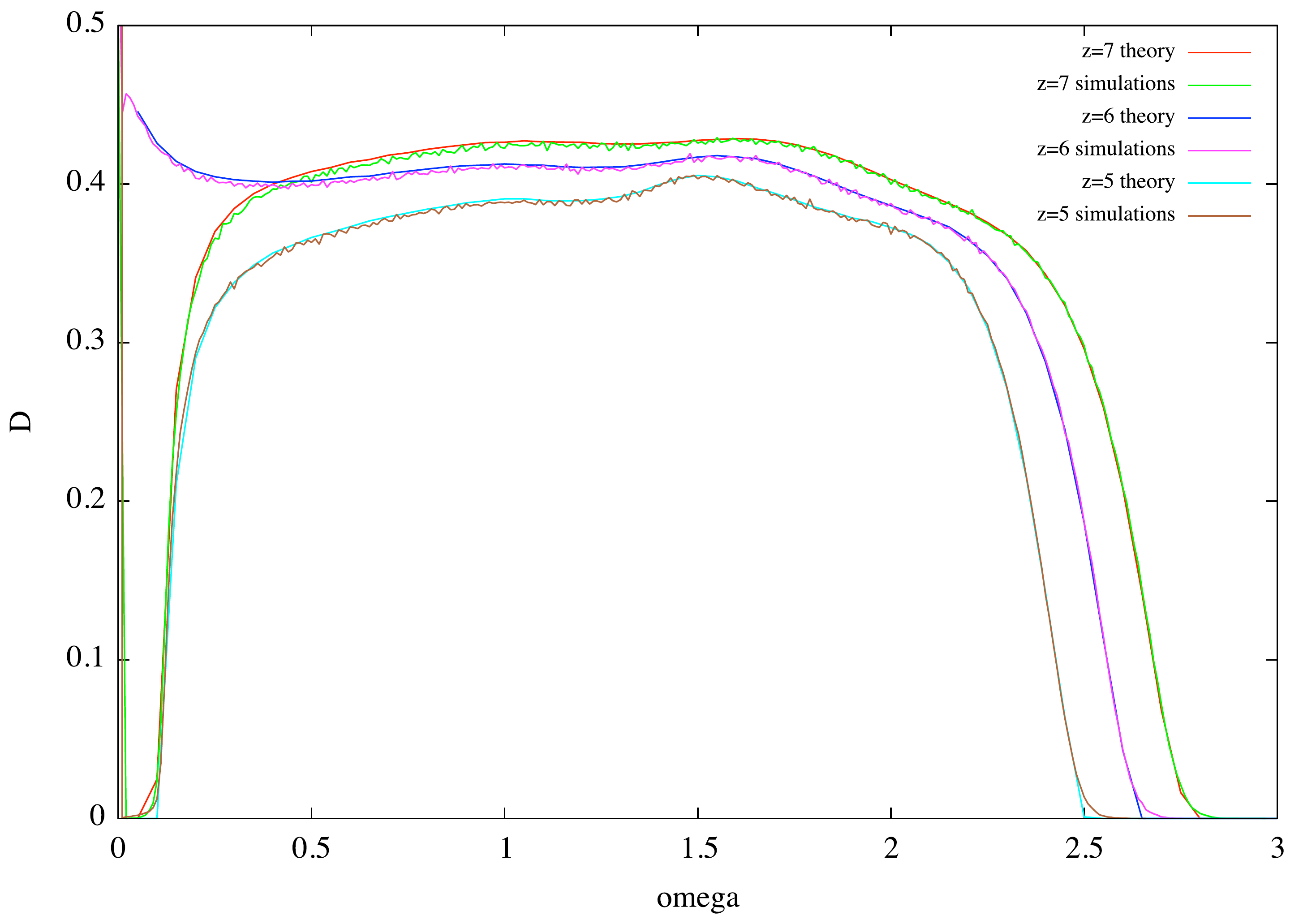}
\end{center}
\caption{
The spectral density ($D(\omega)$) at $\omega$ non-zero for $s=5,6,7$ and D=3. Wiggled lines are the results of  numerical diagonalization of matrices of size $N=200$,
continuous lines analytic results from the cavity method for the lint $N\to\infty$. The agreement is satisfactory (there are some problems in the tails, but it is reasonable for  not too large $N=200)$.
}
\label{fig:paragone}
\end{figure}
The final model  has some similarities with a realistic case: in the real world the local coordination number is fluctuating, the lattice is not Bethe type (i.e. it contains loops) and the vectors $x$ are correlated.
This model can also be studied using an effective medium approximation \cite{EFFECTIVE}. Our aim is to study it analytically and to make a comparison with the numerical simulations at finite $N$.

Let us be precise on the construction we follow of the Bethe lattice. A Bethe lattice is a random locally loopless lattice: there are  $N$ sites and $M$ links that connect different points (there are no links starting and ending on the same point). If the local coordination number $z_i$ (i.e. the number of links starting from  point $i$) if fixed to be equal to $z$, we  have $M=z/2 N$. 
There are two possible constructions, that are equivalent in the large $N$ limit but differs for finite $N$ by $1/N$ corrections. \begin{itemize}
\item
In the usual construction one imposes that two different points are connected by at most one link. In this case we must have $N>z$.
\item No  constraint is present on the number of links going from one point to an other point. In this case the probability of having a double link is $1/N$ and it can be neglected in the limit $N\to\infty$ at fixed $z$. \end{itemize}
We follow the second construction because it allows us to consider also models where $N\leq z$  
\footnote{In constructing the Bethe lattice use the following algorithm (for $N$ even). The sets of the links of our Bethe lattice of coordination $z$ is just the union of those of $z$ random lattices with coordination number equal to 1. The construction of a random lattice with $z=1$ can be trivially done by choosing a random  pair of points, connecting them and eliminating them from the set of points to be connected. }.

\section{The analytic solution of the random spring model}

Usually  models of random matrices on the Bethe Lattice can be solved using the cavity method \cite{CAVITY} that is exact on a Bethe lattices, due to the absence of short loops . One arrives to an equation for the trace of the local resolvent that one solves with population techniques.

This strategy works well also in this case where we can write equations for the cavity one site resolvent $G(\lambda)$ that is  a $D \times D$ matrix \cite{3,5}. The cavity resolvent is the resolvent in a hypotetical problem where only in one point (the cavity) there are $z-1$ neighbours.	

Using the standard procedure \cite{CAVITY} we have the following equation in probability
\begin{equation}
G=_{prob}{1 \over \lambda+ \sum_{k=1,z-1} X_k /\mbox{Tr}(1+G_k X_k)} \, , \label{BASIC}
\end{equation}
where the $X$ are random one-dimensional projectors (i.e. $X_{\mu,\nu}=x_\mu x_\nu$) and the $G_k$'s are random extracted with a probability that eventually must be the same of the probability of the $G$ at the l.h.s of eq. \ref{BASIC}. The equations are well known, also in the matrix case where $D\ne 1$  (see \cite{3,5}). 

The true resolvent $R$ is given by
\begin{equation}
R=_{prob}{1 \over \lambda+ \sum_{k=1,z} X_k/ \mbox{Tr}(1+G_k X_k)}\, .
\end{equation}

In the limit where  goes$\epsilon$ to zero the imaginary part of $\langle Tr(R(\lambda))\rangle|_{\lambda=E-i\epsilon}$ is proportional to the spectral density $\rho(\lambda)$:
\begin{equation}
\lim_{\epsilon \to 0} \langle Tr(R(\lambda))\rangle|_{\lambda=E-i\epsilon} =\pi \rho(E) \, .
\end{equation}  
As soon as $\epsilon$ is not zero (e.g. $10^{-18}$) in a few iterations the result becomes very weakly dependent on $\epsilon$. 

It is evident that the matrix $J$ has rank $zN/2$ and dimensions $ND$ so it must have a $D-z/2$ eigenvalues at $E=0$ in the hypostatic case $z<2 D$. Let us study the isostatic limit by approaching it from the  hypostatic region. 

In the hypostatic region everything is clear. When $z<2 D$ for small $\lambda$ we have
\begin{equation}
\langle R(\lambda)\rangle ={2D-z\over \lambda} +O(1).
\end{equation}

\begin{figure}\begin{center}
\includegraphics[width=.6\textwidth]{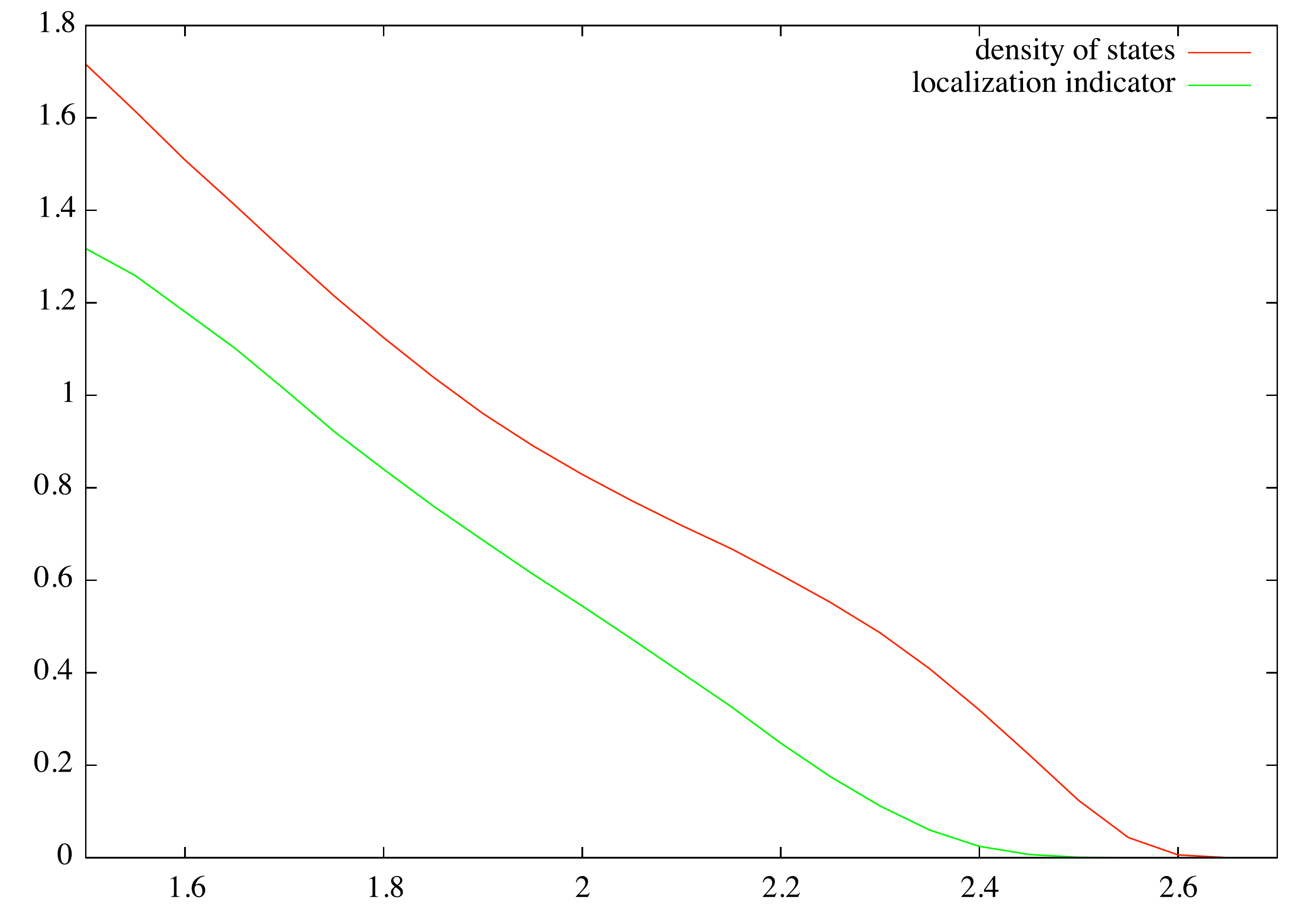}
\end{center}
\caption{
The square of the density of states (red) and the cube of the inverse of the indicator $I_i$ as function of   $\omega$ for $Z=6$ obtained by the cavity method. 
}
\label{LOCO1}
\end{figure}

In this case it is convenient to define
\begin{equation}
\hat G=\lambda G \, .
\end{equation}
In the limit $\lambda \to 0$ the variables $\hat G$ satisfy  the equation
\begin{equation}
\hat G=_{prob}{1 \over 1+ \sum_{k=1,z-1} X_k/ \mbox{Tr}(\hat G_k X_k)} \label{DIS} \, ,
\end{equation}
that should have a solution where
\begin{equation}
\langle \hat R\rangle =2D-z \, .
\end{equation}
The existence of the solution of eq. (\ref{DIS}) for $2D>z$ stems from a simple geometrical fact (i.e. extensive number of zero eigenvalues of $M$) but it is not easy to see the existence of the solution directly from the equations. There should be a simple way to see it,
because it should be valid also in more general Bethe lattice  provided that we do not have sites with the the local coordination number $z_i<D+1$ \footnote{If the local $z_i$ is such that $z_i-1<D$, we may have some troubles in the limit to low frequency because the matrix will be not invertible. On the other hand real world stable configurations of spheres have always $D+1$ contacts so that the condition is natural.}.

Let us consider now the limit $\lambda$ going to zero when $z=2D$. Here no $1/\lambda$ singularity should be present in the $G$, so that we can neglect the $\lambda$ at the denominator. Let us assume that  $G=O(\lambda^{-\gamma})$. In this case we can define
\begin{equation}
\tilde G=\lambda^{\gamma} G\, .
\end{equation}
In the limit $\lambda \to 0$ we get the {\sl homogeneous equation} (i.e. you can rescale all the $\tilde G$ by an arbitrary factor)
\begin{equation}
\tilde G=_{prob}{1 \over  \sum_{k=1,z-1} X_k/ \mbox{Tr}(\tilde G_k X_k)}\label {HOM}\, .
\end{equation}

The same mechanism described in the previous section suggests the natural value of the exponent is $1/2$.
In the numerical solution of the cavity equation   the exponent  $\gamma$ is well compatible with being $1/2$ \footnote{We have checked that the real part is less divergent that the imaginary part near $\omega=0$ and this is possible only if the exponent $\gamma$ is $1/2$.}.
The crucial point is that the disappearance of the solution of eq. (\ref{DIS}) is related to appearance of a solution of eq. (\ref{HOM}): indeed  eq. (\ref{HOM}) is the limit at $z=2D$ of the previous equation (\ref{DIS}) when we normalize the $G$ by a factor $2D-z$.

The final result is that the density of  states at the isostatic point behaves as $\rho(\lambda)\propto \lambda^{-1/2}$ implying that $D(\omega)$ has a finite limit when $\omega$ goes to zero.

\begin{figure}\begin{center}
\includegraphics[width=.6\textwidth]{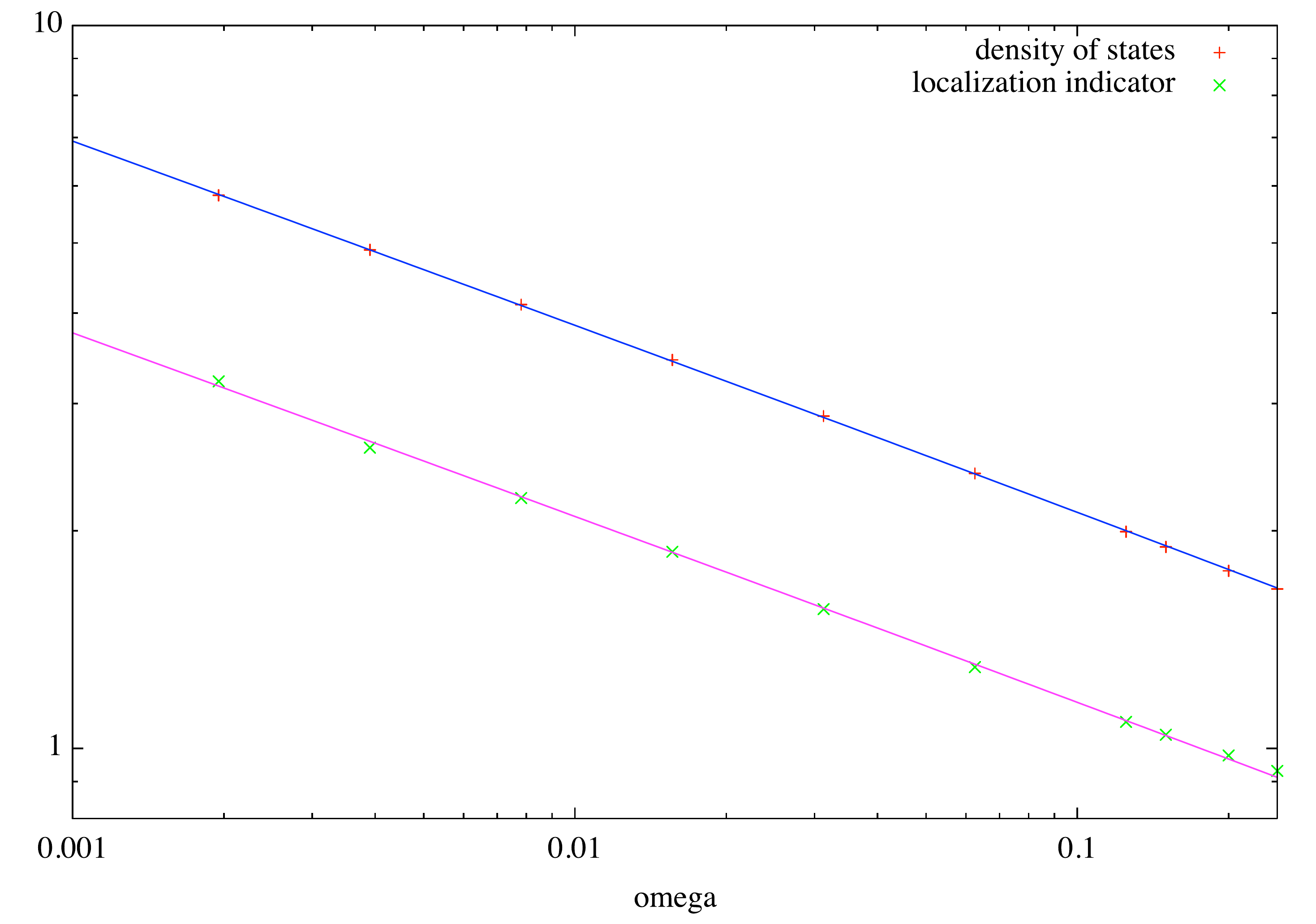}
\end{center}
\caption{
The  density of states ($D(\omega)/\omega)$ to the power 1/4 (red) and the indicator $I_i(\omega)$ as function of   $\omega$ for $Z=6$ obtained by the cavity method (green). The lines are fits assuming a simple $C \omega^{1/4}$ behavior. 
}
\label{LOCO2}
\end{figure}

It may be interesting to note that we can introduce the complex quantity $m\equiv \mbox{Tr}(\tilde G X)$ where $X$ is a random one-dimensional projector.
The recursion equation now becomes
\begin{equation}
m=_{prob}{\mbox{Tr}\left( Y \over \lambda+ \sum_{k=1,z-1} X_k /(1+m_k)\right)},
\end{equation}
where also $Y$ is a random one-dimensional projector.
The advantage of this formulation is that the recursion equation is for the probability of a scalar quantity (that for negative $\lambda$ it is real).

If we call $e_\mu$ the $D$ eigenvalues of $ \sum_{k=1,z-1} X_k /(1+m_k)$, that is a generalized pseudo-Wishart matrix,
we have
\begin{equation}
m=_{prob}{\mbox{Tr}\left( \sum_\mu Y_{\mu,\mu} \over \lambda+ e_\mu\right)}. \label{SCALAR}
\end{equation}
In this formulation it is possible that the integral over the rotational degrees of freedom can be done analytically.

\section{Localization}

For general $z$ ($z\neq z_c$) we expect that the states near the both the lower and the upper edges of the band become localized: the lowest and the highest energy states (excluding the zero modes) are localized: this effect should be similar to standard localization phenomenon as discussed in \cite{3,5} in a similar context. 

At $z= z_c$ something new happens: we find that the region of localized states merges with the disappearing zero modes and the situation is less clear.  

In order to investigate what happens in this particular case we have computed analytically some of the quantities that signal the onset of localization using the solution of the recursion equations.

We can consider various indicators of localization:
\begin{equation}
I_r(\lambda)= \lim_{\epsilon \to 0}{\overline {\left | Re(R_{i,i}(\lambda+i\epsilon))^2\right |} \over \left |\overline {Re(R_{i,i}(\lambda+i\epsilon))}\right | ^2} \, , \ \ \ \ \
I_i(\lambda)= \lim_{\epsilon \to 0}{\overline {\left | Im(R_{i,i}(\lambda+i\epsilon))^2\right |} \over \left |\overline {Im(R_{i,i}(\lambda+i\epsilon))}\right | ^2} \, .
\end{equation}
These quantities are divergent in the localized phase and they are usually finite in extended phase. However they diverge when we approach the localization transition. 

\begin{figure}\begin{center}
\includegraphics[width=.49\textwidth]{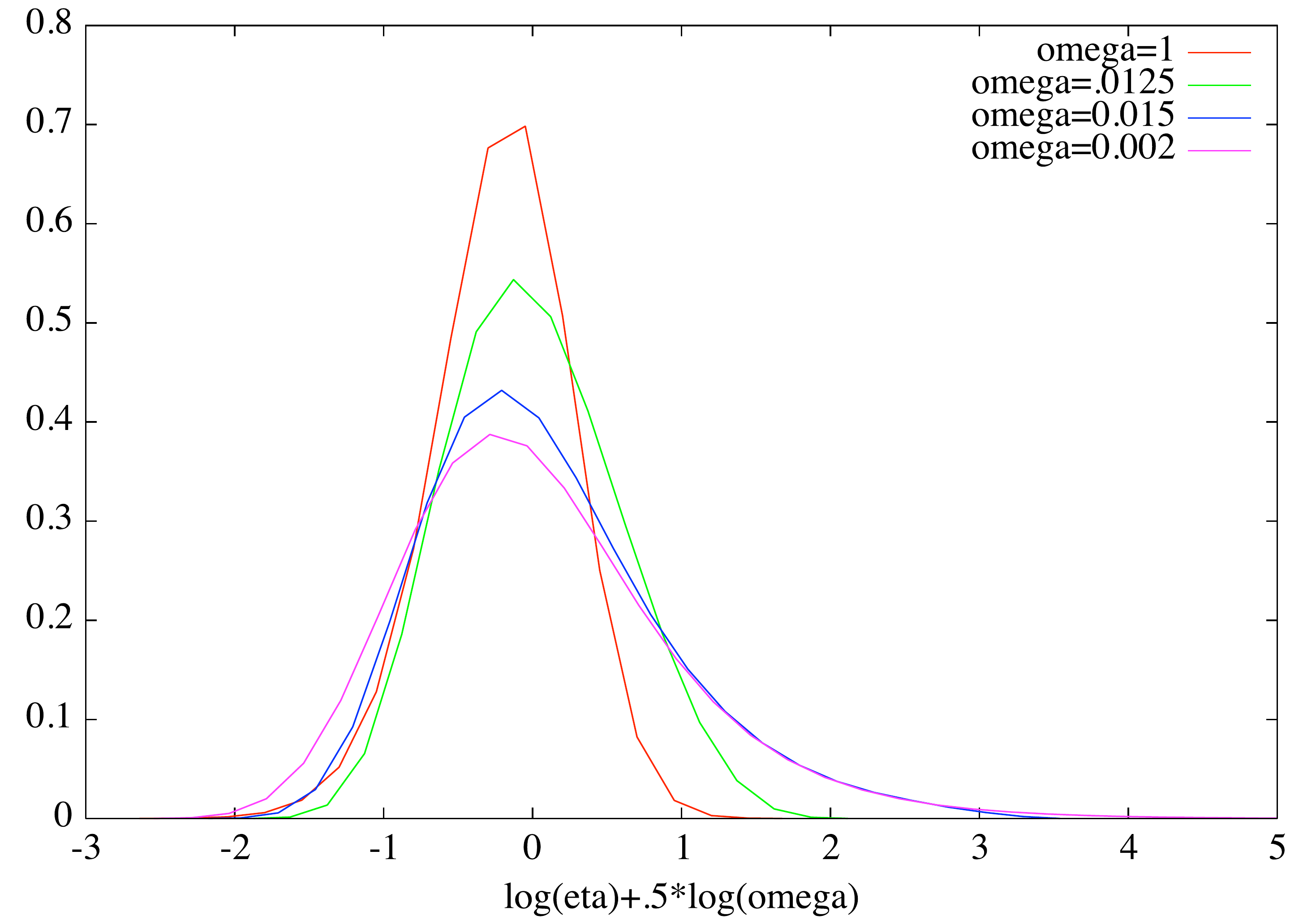} \includegraphics[width=.49\textwidth]{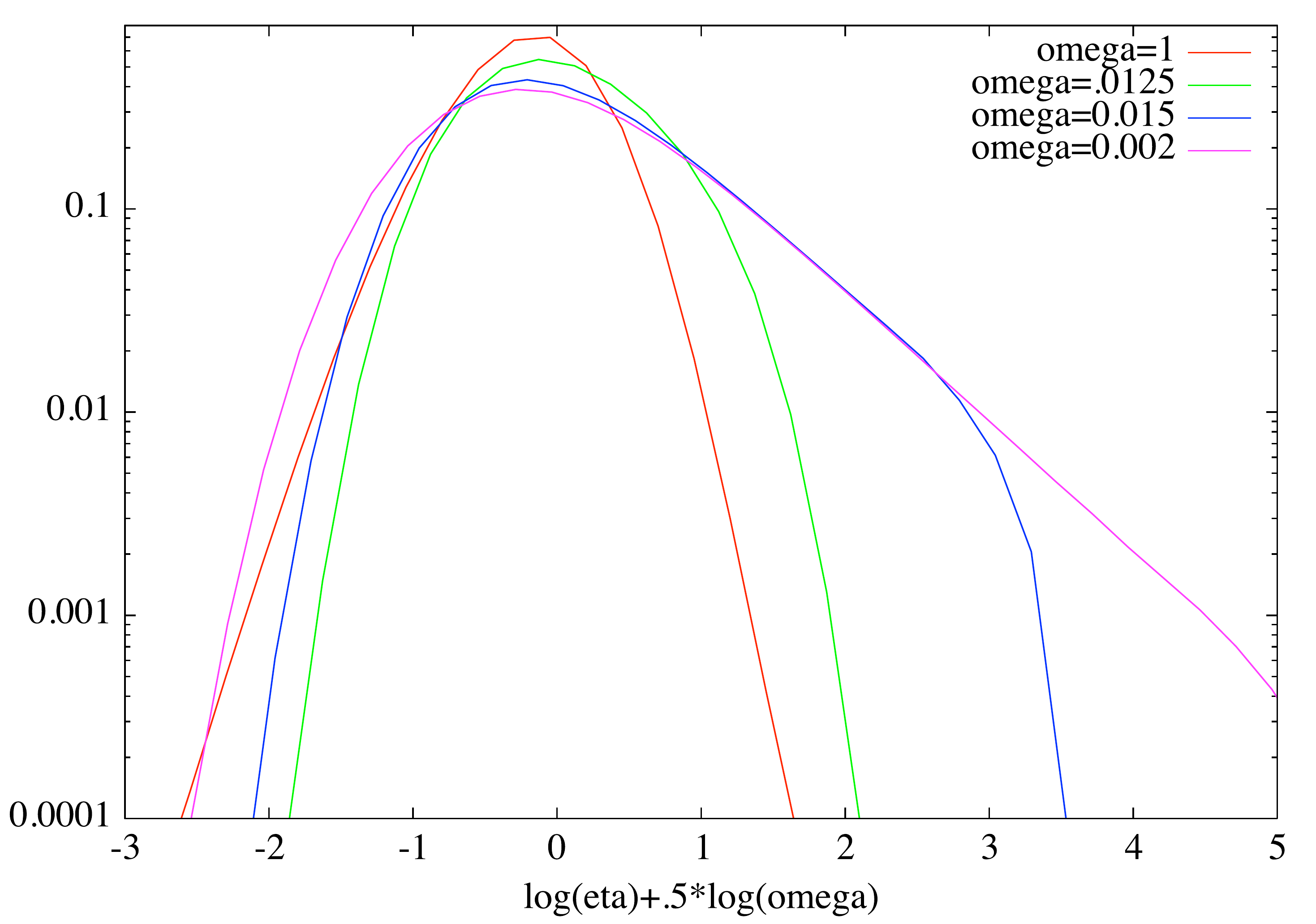}
\end{center}
\caption{
The  curves represent the functions $\eta P(\eta)$, where $\eta$ is the imaginary part of the resolvent in a given point and $P(\eta)$ is the probability function of $\eta$. The values of $\omega $ are $1, 2^{-3}, 2^{-6} , 2^{-9}$. In the left panel we display the function $\eta P(\eta)$ on a linear scale, in the left panel we display the same function on a logarithmic scale.}
\label{LOCO3}
\end{figure}

In fig. (\ref{LOCO2}) we show the behavior  new $\omega=0$ of 
the square of the density of states (red) and the cube of the inverse of the indicator $I_i$ as function of   $\omega$ for $Z=6$ obtained by the cavity method. The edge of the density of states seems to be around $\omega=2.57$ while the localization transition seems to be around $\omega=2.4$. The precise position of the two transition points should be determined by a more careful analysis, taking care of the possibility of exponential singularities in the tail. We  use this data  only to argue that near the right band edge the states are apparently localized although a more careful analysis would be needed in order to reach firm conclusions. This result is  in perfect agreement with the results of \cite{3,5}.

In fig. \ref{LOCO1} we plot  the density of states to the power 1/4 (red) and the indicator $I_i(\omega)$ as function of   $\omega$ for $Z=6$ obtained by the cavity method. The lines are fits assuming a simple $C \omega^{1/4}$ behavior.  The divergence of the indicator at $\omega_1$ is a signal of the avoided localization transition. We notice however that the increase of the indicator $I_i(\omega)$ is rather modest (a factor 3) so that  it quite difficult to give a reliable estimate the power of $\omega$. 

The effect of the incipient localization can be seen quite dramatically if we plot the probability distribution of the imaginary part of the resolvent on a logarithmically scale.  This is done in fig. (\ref{LOCO3}), where  we plot the probability distribution of the local imaginary part of the resolvent; the values of $\omega $ are $1, 2^{-3}, 2^{-6} , 2^{-9}$. One clearly sees that there is a widening	of the probability distribution in a logarithmic scale. However the behavior is somewhat complex and we have not attempted to derive scaling formulae for this quantity. The numerical solution of the population equation becomes more and more difficult when we approach the localization transition (the effects of a finite size of the population become more and more important, so that one need to use very large population samples at small $\omega$).

\begin{figure}\begin{center}
\includegraphics[width=.6\textwidth]{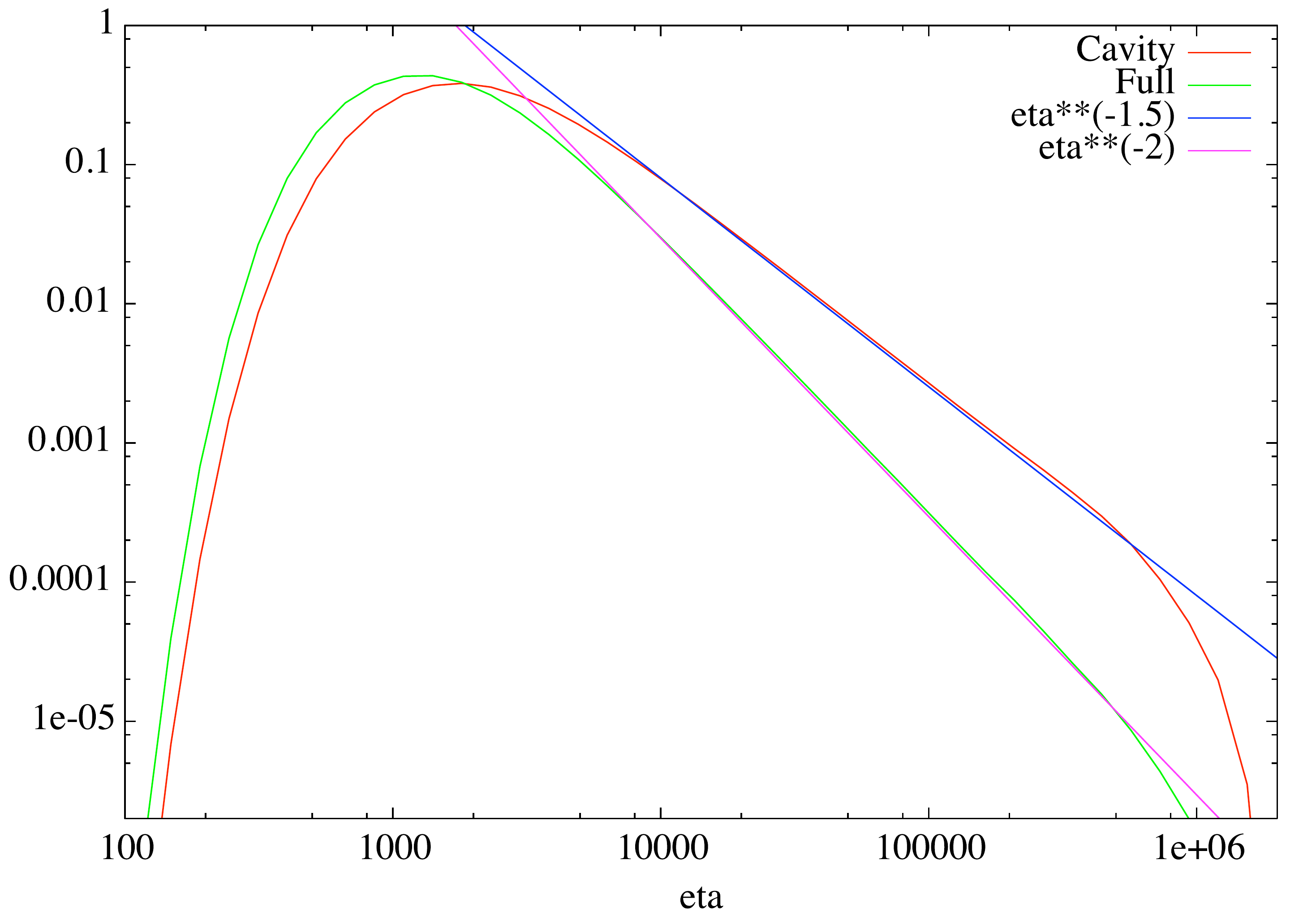}
\end{center}
\caption{
We plot the probability distribution of the imaginary parts of the resolvent at $\omega=10^{-3}$The  green curve is the function $\eta P(\eta)$, where $\eta$ is the imaginary part of the resolvent in a given point and $P(\eta)$ is the probability function of $\eta$. The red curve is the probability distribution of the imaginary part of the cavity resolvent. The two straigth lines are  fits assuming a simple $C \eta^{-2}$ and $C' \eta^{-3/2}$  behavior. }
\label{LOCO4}
\end{figure}

The appearance of a tail at large values of the imaginary part really stands out. We have looked more carefully to this region in fig. (\ref{LOCO4}). The green curve is the function $\eta P(\eta)$, where $\eta$ is the imaginary part of the resolvent in a given point and $P(\eta)$ is the probability function of $\eta$. The red curve is the probability distribution of the imaginary part of the cavity resolvent. The two straight lines are  fits assuming a simple $C \eta^{-2}$ and $C' \eta^{-3/2}$  behavior.  These tails are related to rare cases where the $D$ ($D-1$) vector $X$ are nearly coplanar, so that the eigenvalues of $R$ may be as large a $\lambda$. The values of the exponents can be readily understood with a simple argument \footnote{We have to consider the probability distribution of the smallest eigenvalue $\mu$ of a Wishart matrix. One finds that in dimensions $N$ the probability distribution of $\mu$ for small $\mu$ is given by $W(\mu)\propto \mu^{-1+(M+1-N)/2}$. In our case eq. (\ref{SCALAR}) corresponds to $N=3$ and to $M=6$   ($M=5$) for the resolvent (the cavity resolvent): the imaginary part of the  resolvent ($\eta$) is proportional to $\mu^{-1}$ and this way we obtain the needed result. }. 

The presence of these fat tails is related to the existence of quasi-localised states. One finds that
\begin{equation}
P(\eta) \propto \eta^{-3}\, ,
\end{equation}
where $P(\eta)$ is the probability distribution of $\eta$. i.e. the imaginary part of the resolvent in a given point. One finds that the contribution from these only this tail diverge logarithmically with $\omega$. A power divergence in $I_i(\omega)$ for small $\omega$ can be explained only by the widening of $P(\eta)$ near the peak. Separating a small power of $\omega$ from a  logarithmic divergence is not an easy job. It would be interesting to understand what happens in dimensions $D$ greater or smaller than $D$.

\section{The large $D$ limit}
Although we we would like o understand the limit $N$ to infinity at fixed $D$ it is interesting to consider the case of large $D$ and $z$ where analytic computations can be done in a closed way.

The limit $D\to\infty$ has been investigated in the case of hard spheres. It has been found that many interesting properties of hard sphere have a weak dependence on the dimension $D$ \cite{LARGED0} and that many analytic computations are possible in that limit \cite{LARGED1}. It is remarkable that the various exponents connecting various	quantities in the jamming limit can be computed analytically \cite{FINAL}: e.g. the radius of the cages goes to zero at large pressure as $p^{-\kappa}$ with $\kappa=1.41574(1)$.

We consider the limit $D\to\infty$ at fixed $\gamma=z/D$.  The first case we can consider is $N=2$.
In this case it is convenient to define  $s_{\pm}=\delta_1\pm \delta_2$. We find that the matrix $J$ is zero on the $+$ space and it reduces to the Wishart model in the $-$ space. The same computation for $N=3$ is more complex.

To make further progresses we find  convenient to consider first the $N\to\infty$ limit and later the $D\to\infty$ limit.
In this way we control the model in the limit $D=\infty$.

We can  do an explicit computation for large $D$. In this case the law of large numbers imply that $m$ does not fluctuate, we can set  $Y_{\mu,\mu}$ to its average and the previous equation (\ref{SCALAR}) becomes
\begin{equation}
m=D^{-1}{\mbox{Tr}\left( 1 \over \lambda+ W(z-1,D)/(1+m)\right)}=
D^{-1}{\mbox{Tr}\left(1+ m \over (1+m) \lambda+ W(z-1,D)\right)},
\end{equation}
where $W(2D-1,D)$ is a pseudo-Wishart matrix. In the limit of $D\to \infty$ we get
\begin{equation}
s=-s q(-s\lambda,\gamma)+1 \, ,
\end{equation}
where $s=1+m$.

In the limit of $s \lambda$  small we can use the relation $q(x,\alpha)=1/(1-\alpha)-A(\alpha) x^2+O(x^3)$, with $A(\alpha)=\alpha/(\alpha-1)^3$.  We get
\begin{equation}
s=s/(\gamma-1)+A(\gamma) s^2\lambda+1+O(s^3)
\end{equation}
that for $\gamma=2$ gives $s\propto \lambda^{-1/2}$ as it should be. 

Also in this case the spectrum is of Marcenko-Pastur type. Therefore in the $D\to \infty$ limit we recover the same picture as in the Wishart case. The $\lambda^{-1/2}$ singularity in the spectrum seems to be a quite robust result, that quite likely survives also in a  $1/D$ expansion

\section{Conclusions}

Where do we stand?  The Wishart model captures most of the physics of the problem. In particular for $\alpha>\alpha_c$ the gap $\lambda_c(\alpha)$ is proportional to $ (\alpha-\alpha_c)^2$. This correspond to $\omega_c\propto \alpha-\alpha_c$, that compares well with the numerical results for the DD model of \cite{DD}, where the data are well fitted by $\omega_c\propto (z-z_{c})^/nu$ with $\nu=.94$.  

The new phenomenon (present only in models with a finite coordination number) is the quasilocalization near $\omega=0$. This phenomenon can be observed numerically if we consider the resolvent  for negative values of $\lambda$ where it is real.  

At this end it may be convenient to define a resolvent projected in the direction orthogonal to that of the zero modes due translational invariance. The projected resolvent $Q$ satisfies the equation
\begin{equation}
\lambda Q(\lambda)_{i,j}-\sum_k J_{i,k}Q(\lambda)_{k,j}=\delta_{i,j}-{1\over N}\, .
\end{equation}
In other words $Q(\lambda)$ is the response to a monopole force (this is similar to the dipole case considered in \cite {W}). The difference beween $Q$ and $R$ is vanishes for infinite $N$, however it may be very important at finite $N$ and very small $\lambda$.

The behaviour of $Q_{i,i}(\lambda)$ for small negative $\lambda$ should be similar to the  behaviour of $Im(R(\lambda))$ for small positive $\lambda$. Large fluctuations in $Q_{i,i}(\lambda)$ are expected and this is likely related to the fluctuations of the response to a dipole force observed in  \cite {W}.

It is clear that further work is needed to go beyond mean field theory and to arrive to definite predictions for finite dimensional harmonic spheres packing.  It is possible that the ideas developed in \cite{W,NATURE,RMFIM} may be useful in this context.

\section*{Acknowledgements} I am very grateful to Pierfrancesco Urbani and Francesco Zamponi for many inspiring discussions and for a careful reading of the manuscript. I would also like to thank 
Silvio Franz and Andrea Liu  illuminating discussions.
Financial support was provided by
the European Research Council through ERC grant agreement
no. 247328.

\end{document}